\documentclass[referee]{aa}
\usepackage{graphicx}
\usepackage{amssymb}
\usepackage{natbib}
\bibpunct{(}{)}{;}{a}{}{,} 

\begin{document}

\title{Birth rates of SGRs and AXPs: delayed amplification of magnetic field}

\author{Denis Leahy and Rachid Ouyed}

\institute{Department of Physics and Astronomy, University of Calgary, 
2500 University Drive NW, Calgary, Alberta, T2N 1N4 Canada\thanks{email:leahy@iras.ucalgary.ca; ouyed@phas.ucalgary.ca}}

\date{Received <date>; accepted <date> }

\authorrunning{Leahy\&Ouyed}

\titlerunning{SGRs/AXPs birthrate}

\abstract{
We present new determination of the birth rate of AXPs and SGRS and their
 associated SNRs. We find a high birth rate of $1/(500\ {\rm yr})$ to $1/(1000\ {\rm yr})$  for AXPs/SGRs 
  and their associated SNRs. These high rates suggest that all massive stars (greater
    than $\sim (23$-$32) M_{\odot}$) give rise to remnants with magnetar-like fields.
     Observations indicate a limited fraction of high magnetic fields in these
     progenitors thus our study necessarily implies  magnetic field amplification.
      Dynamo mechanisms
       during the birth  of the neutron stars require spin rates much faster than
        either observations or theory indicate.
    Here, we propose that neutron stars form with normal ($\sim 10^{12}$ G) magnetic
    fields, which are then amplified to $10^{14}$-$10^{15}$ G after a delay
     of  hundreds of years. The  amplification is speculated to be a consequence of color ferromagnetism and to occur  after  the neutron star core reaches quark-deconfinement density.  This delayed amplification alleviates many  difficulties in interpreting simultaneously the high birth rate,  high 
        magnetic fields, and state of isolation of AXPs/SGRs and their link to massive stars.
\keywords{stars: evolution --- stars: neutron: SGRs/AXPs --- supernovae: SNR} }

\maketitle

\section{Introduction}

Early studies of association of Anomalous X-ray Pulsars (AXPs) with  supernova remnants (SNRs) suggested that  5\% of core-collapse SN results in AXPs (Gaensler et al. 1999). 
 This was based on 3 SNR associations out of a total of 6 AXPs. 
  Since then evidence has mounted that AXPs and soft gamma-ray repeaters (SGRs)
   are the same type of objects (Gavriil et al. 2002) and more AXPs, 
     SGRs and associated SNRs have been identified. Thus it it timely to revisit
      the issue of AXPs/SGRs birthrates. 
     
     In this study  we present an updated investigation of the birth rate of AXPs/SGRs
      and in addition, for the first time,  the birth rate of associated SNRs is given. 
       Since AXPs/SGRs ages rely on spin-down age estimates whereas SNRs
        ages are based  on shock expansion models,  this constitutes two independent estimates
          for birth rates.    Both samples yield a high birth rate for AXPs/SGRs\footnote{An independent
           study by Gill\&Heyl (2007), based on  a population
synthesis of AXPs detected in the ROSAT All-Sky Survey, yields a birth rate of $\sim 0.22$ per
           century.} of
           $(1/5)$-$(1/10)$ of all core-collapse SNe, higher than previously appreciated.
            This high frequency of occurrence of AXPs/SGRs brings into focus
              issues related to the origin of the strong magnetic fields which we address here.
  This paper is presented as follows: \S 2 describes the methods and presents the
 birth rate results, and \S 3  discusses the implications. Our model, based
  on a delayed amplification of magnetic field, is presented in \S 4 before 
  concluding in \S 5.

\section{Birth rate based on spin-down and SNR ages}
 
 One can derive birth rates by fitting a linear trend to the observed cumulative
  number versus age relation. For SNRs, we take data from Marsden et al. (2001)
    and for spin-down ages we used $P$ and $\dot{P}$ from the Australia Telescope
     National Facility (ATNF) website.
     We supplemented the ATNF data with recent updates from the literature (see Table 1).
     The sample consists of 5 SGRs and 10 AXPs and 9 associations with SNRs.
      Of these objects AXP1E2259$+$586 and AXP4U0142$+$615 were omitted
      from our sample since these may have disks (Ertan et al. 2006; Gonzalez et al. 2007) which make the spin-down
       age unreliable.   We note that including these objects
        did not change our estimates of birth rate from spin-down ages
         but resulted in worse fits.  For the SNR age we adopted the geometric mean of the lower
   and upper ages given in Marsden et al. (2001).

\begin{table*}[t!]
 \caption{{\small SGRs/AXPs data  from
 Marsden et al. (2001), the atnf pulsar catalogue (Manchester et al. 2005),
  and Camilo (2007).  Ages  in years.}}
\begin{center}
\begin{tabular}{|c|c|c|c|c|}\hline
 Source & $\tau_{\rm SD}$ & $\tau_{\rm SNR, lower}$ & $\tau_{\rm SNR, upper}$ & $\tau_{\rm SNR, mean}$\\ \hline
 SGR1806$-$20 &  281 & 3500 & 30000 & 10300\\
SGR1900$+$14 & 1050 & 9600 & 30000 & 17000\\
SGR0525$-$66  & 1960 & 5000 & 16000 & 8940\\
SGR1627$-$41  & N/A & 2600 & 30000 & 8830\\
SGR1801$-$23 & N/A & 2400 & 30000 & 8490\\
AXP1E1048$-$5937 & 2680 & 9800 & 30000 & 17200\\
AXP1E1841$-$045 & 4510 & 500 & 2500 & 1120\\
AXP1845$-$0258 & N/A & 600 & 30000 & 4240\\
AXPRXS1708$-$4009 & 8960 & N/A & N/A&  N/A \\
TAXPXTEJ18010$-$197 & 4260 & N/A & N/A & N/A \\
TAXPJ0100$-$7211 & 6760 & N/A & N/A & N/A \\
AXP1547$-$5408 & 1400 & N/A & N/A & N/A \\\hline
AXP1E2259$+$586 & 228000 & 3000 & 17000 & 7140\\
AXP4U0142$+$615 & 70200 & N/A & N/A&  N/A \\\hline
\end{tabular}
\end{center}
\label{default}
\end{table*}%
   
  The left panel of Figure \ref{fig:birthsdsnr} shows the cumulative number of
  associated SNRs ($N_{\rm SNR}$, diamonds and dot-dashed line) versus SNR age. To show the
 uncertainties in the ages we also plot the minimum and maximum ages for each SNR.
  The solid line is the expected number versus age relation for a constant
  birth rate of $1/(1700\ {\rm yr})$. In the right panel we show the cumulative number of
  SGRs/AXPs  ($N_{\rm SD}$, circles and dashed line) versus spin-down age.
    The solid line is the expected number versus age relation for a constant
  birth rate of $1/(500\ {\rm yr})$. In the right panel we re-plot the cumulative
   number versus age relation for associated SNRs scaled up by a factor of 3 ($3 N_{\rm SNR}$,
   diamonds and dotted line); the dot-dashed line re-plots the SNR birth rate of $1/(1700\ {\rm yr})$
     from the left panel.

   We fit the data by a constant birth rate model:   one set
   of fits assumes normal statistics and another set uses a 
   robust estimator (e.g. \S 14 in Press et al. 1989). The robust estimator
   uses the sum of absolute values of differences rather than the sum of squares
    and thus gives less weight to outliers.  The results assuming normal statistics give an SGR/AXP
   birth rate from spin-down of $1/(500\ {\rm yr})$ with
    a $1\sigma$ range of  $1/(400\ {\rm yr})$ to $1/(570\ {\rm yr})$,
    assuming a 50\% uncertainty in spin-down age. For the robust
    estimator the best fit  is $1/(525\ {\rm yr})$ consistent with the above.
       For the associated SNRs, the resulting 
   birth rate from the normal estimator is $1/(1600\ {\rm yr})$ with a $1\sigma$ range of  $1/(1500\ {\rm yr})$ to $1/(1770\ {\rm yr})$ while the  robust estimator gave $1/(1770\ {\rm yr})$.  
     In all cases the $\chi^2$ values
    were acceptable indicating a constant birth rate fit is an acceptable model.

\subsection{Birth rate comparison}

    The  birth rate derived from associated SNRs
     is $\sim 1/3$ of the birth rate derived from spin-down ages.
     There are two effects that could account for such
     a discrepancy. 
     
\begin{figure*}[t!]
\centerline{\includegraphics[width=0.5\textwidth,angle=0]{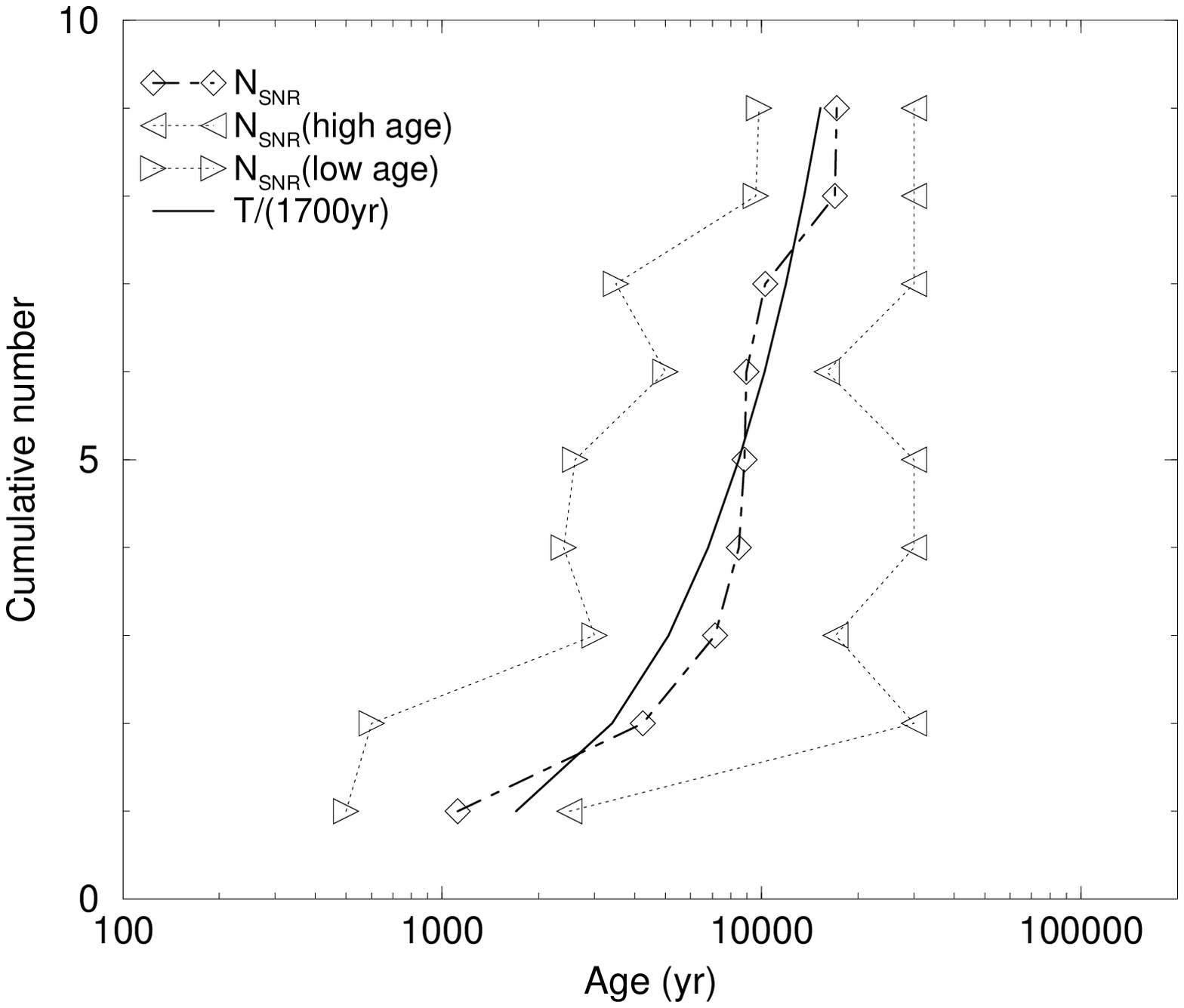}
\includegraphics[width=0.5\textwidth,angle=0]{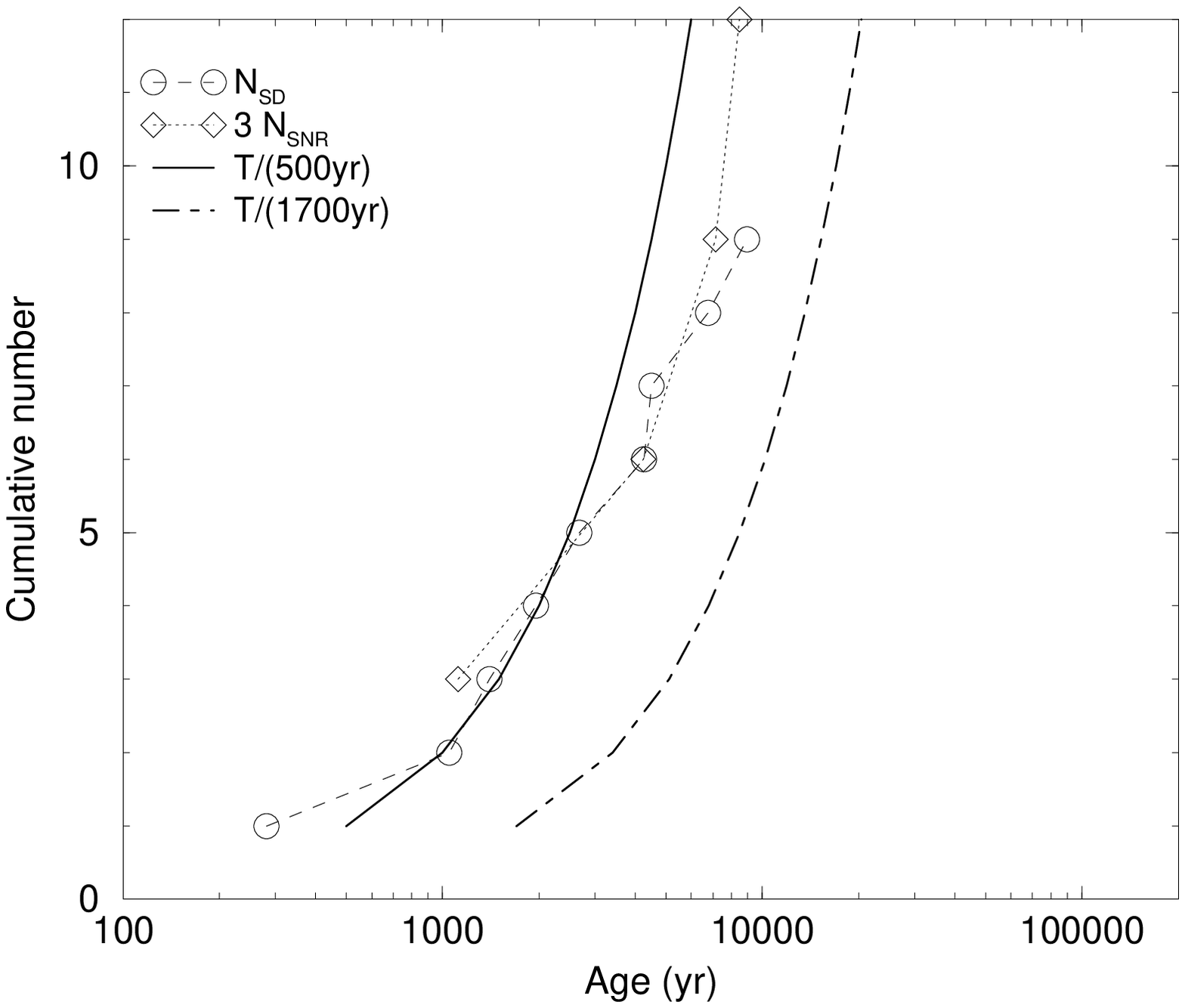}} 
\caption{\label{fig:birthsdsnr}
 The left panel shows the cumulative age distribution for SNRs associated with AXPs and SGRs (diamonds and dot-dashed line).  The solid line is the expected distribution for constant birth rate  of $1/(1700\ {\rm  yr})$.
  The triangles indicate the upper and lower SNR age limits.  The right panel shows the cumulative age distribution for AXPs/SGRs (circles and dashed line). The solid line is the expected distribution for constant birth rate  of $1/(500\ {\rm  yr})$. The diamonds and dotted line is the cumulative
  distribution for associated SNRs (from left panel) scaled up by a factor of 3. To better illustrate
   this scaling, the $1/(1700\ {\rm  yr})$ line  is re-plotted (dot-dashed line).
 }
\end{figure*}

     One effect is incompleteness of either sample,
      which would increase the birth rate of that sample; in this case
       incompleteness of the SNR sample could increase the birth rate
       to match the birth rate from spin-down.  As can be seen from the right panel of Figure
     \ref{fig:birthsdsnr}, if we increase the number of SNR by a
     factor of about 3 we obtain good agreement with the 
      number versus age relation for AXPs/SGRs.  It is worth pointing out  that  of
         the 15 AXPs/SGRs, 9 show associated SNRs. Since all
          AXPs/SGRs have been searched for associated SNRs,
           the SNRs are too faint to be seen. This is either due to:
           (i) the SNR is old; (ii) the SNR is not detected due to confusion;
            (iii) the SNR is young but the environment has low density.
            Either of the latter two situations 
           suggests incompleteness, with a factor of about  $\sim 15/9$,
            raising the birth rate estimate from associated SNRs to $\sim 1/(1000\ {\rm yr})$;
           this is not  enough to account for the difference in birth rates.
           For SGRs/AXPs birth rate if there is incompleteness in the sample
            then the birth rate increases above $1/(500\ {\rm yr})$. However,
             these objects are fairly bright in X-rays so only transient
             SGRs/AXPs would contribute to incompleteness. The high
             birth rate we derived indicate that there cannot be very many transients.
              Thus the  incompleteness cannot be an important factor otherwise
               we overproduce AXPs/SGRs compared to the total SN rate in the Galaxy.
               
           The second effect is that SNRs or AXPs/SGRs ages could be systematically off by
            a factor of $\sim 2$. 
            In effect instead of shifting points vertically in Figure
        \ref{fig:birthsdsnr}, the points are shifted horizontally.
         There is no reason why the SNR ages should be 
        systematically too large by up to a factor of $\sim 2$. 
        However there are reasons
        to believe that spin-down ages may systematically be off.  
         The general spin-down formula for braking index $n$, $\dot{\Omega}= - K\Omega^{n}$ 
         (where $K$ is a constant; e.g. M\'esz\'aros1992), implies a spin-down age $\tau = P/((n-1)\dot{P})$.
         Table 1 assumes the vacuum dipole case with $n=3$. However the few pulsars
          with measured  braking indices have values $n >2$ with the exception of the
          Vela pulsar with $n=1.4$ (Lyne et al. 1996). For $n=2$ the spin-down age
          is twice that listed in Table 1: this can bring the spin-down derived birth rate
 down to $\sim 1/(1000\ {\rm yr})$   in agreement with the SNR derived value corrected
 for incompleteness. 
 
 However,  for 4 of the objects listed
          (SGR1806$-$20, SGR1900$+$14, SGR0525$-$66 and, AXP1E1048$-$5937)
          a doubled spin-down age is still not enough to remove the discrepancy
           between spin-down age and the lower limit to the SNR age.      
        If the initial period of the neutron star is below $\sim 1$ ms,
       the moment of inertia decreases as it spins down (e.g. Berti\&Stergioulas, 2004). The spin-down formula given above assumes constant moment of inertia
        and thus a constant $K$.  Taking into account changes in the oblateness (moment of inertia)
as the star spin-down from millisecond period leads to no more than  20-30\% 
 increase in age estimate. This is not large enough to explain the discrepancy.
      On the other hand, spin-down ages assume constant magnetic field. Including
magnetic field decay will decrease these ages and worsen the discrepancy.

        To summarize this section, the AXPs/SGRs birth rate  is about
         $1/(500\ {\rm yr})$ ($n=3$) to $1/(1000\ {\rm yr})$ ($n=2$). The latter is
         consistent with the SNR-derived rate corrected for incompleteness.
          We still need to explain a large discrepancy in age for the 4
          cases mentioned above.  We suggest a  time delay from SNR explosion to the onset of spin-down
           to explain these cases (see section \ref{sec:delay}).

\section{Implications}\label{sec:implication}

A birth rate of  $1/(500\ {\rm yr})$ to $1/(1000\ {\rm yr})$  implies $1/5$ to $1/10$ of all core-collapse
     SNe lead to SGRs/AXPs. To interpret this we use 
      a Scalo mass function,  minimum and maximum SN progenitor masses of $9M_{\odot}$
       and $60M_{\odot}$.       
      For the $1/5$ case, the SGR/AXP progenitor mass range is $23M_{\odot}$ to $60M_{\odot}$;
        for the $1/10$ case, the SGR/AXP progenitor mass range is $32M_{\odot}$ to $60M_{\odot}$.
         The ranges can be shifted as long as they give the same fraction of SNe
         that lead to SGRs/AXPs (e.g. $20M_{\odot}$ to $40M_{\odot}$ for the $1/5$
         case). 
      
       An alternate possibility is that  $1/5$ to $1/10$ of
    SNe for all progenitor masses produce AXPs/SGRs. However
      observations of associated SNRs indicate that SGRs/AXPs are associated with massive star
      progenitors (Gaensler 1999) so we favor 100\% production at the high-mass end with
       $M > M_{\rm low}=(23$-$32)M_{\odot}$.

This raises the following questions:

\begin{enumerate}
\item How  do all progenitors  with $M\ge M_{\rm low}$ generate
 $>10^{14}$ G fields in their compact remnants?  
 \item Why is there a sudden jump in the magnetic field strength between
  compact remnants from progenitors with mass greater than $ M_{\rm low}$ (i.e.
   $B\sim 10^{14}$ G)
   and those with mass less  than $M_{\rm low}$ ($B\sim 10^{12}$ G).
   \item Why are all compact remnants from $M\ge M_{\rm low}$ progenitors
    isolated whereas the progenitors show a high binary fraction?
\end{enumerate}

In regards to point 1 above, observations of OB stars (Petit et al. 2007) found 
 3 out of 8 with $\sim$ kG fields and one out of two massive stars
 with $\sim$ kG fields. Despite the paucity of data, this indicates
  that not all massive stars are strongly magnetic. 
   The fossil field hypothesis (Ferrario \& Wickramasinghe
2006) predicts even lower numbers of magnetic massive stars than observed.
 Our study implies that a magnetic field amplification mechanism
  is required to explain high fields in all compact remnants from
  massive stars.  One natural mechanism would be dynamo generation during
   neutron star formation (Thompson\&Duncan 1993).
  However as shown by Vink\&Kuiper (2006) the SNRs associated with SGRs
  and AXPs  have normal explosion energy ($\sim 10^{51}$ erg) conservatively limiting the
   birth periods to $> 5$ ms. This provides a major challenge for the dynamo mechanism
   for the generation of SGR/AXP magnetic field strengths.
       Heger et al. (2005) also consider the spin periods
    of neutron stars at birth from  massive stars using
     a stellar evolution code. They calculate the evolution of 12-35 $M_{\odot}$
    progenitors  including magnetic
   field and angular momentum transport.  Their Table 4, gives results of
   $\sim$ 15 ms (for $12M_{\odot}$) to 3 ms (for $35M_{\odot}$),  many times slower than previously obtained in calculations   ignoring magnetic torques. This in effect also argues against 
    the dynamo mechanism, which requires sub-ms periods,
     to generate magnetar-like fields for stars of $35M_{\odot}$ or less.
   This leaves us with the dilemma of how to account for the
   strong magnetic fields inferred for AXPs/SGRs (i.e. all descendants of stars
    more massive than $M_{\rm low}$)?

 Alpar (2001) instead suggests that all AXPs/SGRs have normal magnetic fields ($\sim 10^{12}$ G)\footnote{An alternate proposal by Dar\&DeR\'ujula (2000) involving
 normal magnetic field strength suggest a conversion to quark matter accompanied by 
  a slow gravitational contraction to power the observed emission.}.
  The large  spin-down rates of  AXPs/SGRs are then explained by accretion (with propeller mechanism
   to give the large positive $\dot{P}$; Chatterjee et al. 2000) from a fall-back disk following
  the SN explosion. This would avoid the problem of generating strong magnetic fields 
   in all stars with mass $>M_{\rm low}$.  
      Wang et al. (2006) find support for  a debris disk around 4U0142$+$61
     thus  several destruction mechanisms such as radiation, magnetic propeller and flares 
     which could limit the  lifetime of fall-back disks, at least in this case 
      are not effective.  However,  debris-disk models have difficulties explaining
    the large negative $\dot{P}$  (i.e. spin-up) occasionally observed in AXPs/SGRs.

 Point 2 suggests some new physical mechanism for magnetic
 field amplification that sets in, independent of progenitor magnetic
 field, but dependent on progenitor mass.  Finally, for point 3, the
  explosion of the progenitor in many cases leads to binary disruption.
   However, there still should remain a  fraction of binary
    remnants. In our model, we suggest the second explosion (see below) further
    reduces the binary fraction.

\section{Proposed explanation}\label{sec:delay}

We offer an alternate explanation which allows normal
magnetic fields for neutron stars born from progenitors with mass $>M_{\rm low}$,
 in addition to lower mass progenitors.  In this picture, the magnetic field
amplification occurs long after the neutron star formation, but
only for neutron stars born from massive progenitors.  The amplification
  occurs during the conversion from baryonic matter
  to quark matter which happens after the neutron star core
  reaches quark deconfinement density.

  Amplification of the magnetic field up to $10^{15}$ G
  can be achieved as a result  of color-ferromagnetism (Iwazaki 2005) during the
    phase transition.
   The magnetization here is unlike the case of a normal ferromagnet
where spontaneous magnetization occurs as the temperature falls below 
the Curie temperature (there the order
parameter is  the spontaneous magnetization $M$ namely, the expectation value of spin over the sample, and an external field is required to  impose 
domain alignment). Color-ferromagnetism instead is dictated by the Savvidy effect which
 is an instability of the vacuum due to infrared singularities (Savvidy 1977).
 In Color-Ferromagnetic quark matter (SU(2)), the color magnetic field is generated spontaneously not by alignment
 of quark color spins, but by the dynamics of the gluons (Iwazaki et al. 2005). 
Due to the nature of fractional quantum hall states, the color magnetic field can
exist globally in the quark matter, without domain structure. This
 global uniform field is the minimum energy state (Iwazaki 2005 and references therein).

   Staff et al. (2006) discuss the time delay from neutron star formation
   to deconfinement in the core and subsequent quark star formation (which
   occurs in an explosive manner, called a Quark-Nova or QN; Ouyed et al. 2002;
    Ker\"anen\&Ouyed 2003; Ker\"anen et al. 2005).
    They found that neutron stars with, (i) mass greater than $\sim 1.5M_{\odot}$, (ii) initial
    periods less than $\sim 3$ ms and, (iii) magnetic fields less than $\sim 10^{12}$ G, experience
    deconfinement after several hundred years (see Table 2 and discussion
    in Staff et al. 2006).  These numbers are interesting as they imply
      progenitors consistent with those we discussed above in the
      context of birth rates of AXPs/SGRs (i.e. $M>M_{\rm low}$, 
       $B\sim 10^{12}$ G and, periods of a few milliseconds). For
        more massive neutron stars (with progenitors mass around approximately  $\sim$ 50-60$M_{\odot}$) the delay
         is days rather than centuries leading to an energized SN instead (Leahy\&Ouyed
         2007). More massive progenitors lead to black holes.

 To represent the idea of a delayed amplification of the magnetic
 field, we write the time since SN explosion as
 \begin{equation}\label{eq:one}
 \tau_{\rm SNR} = \tau_{NS}+\tau_{\rm QS}\ ,
 \end{equation}
 where $\tau_{\rm NS}$ and $\tau_{\rm QS}$ are the time the compact
 object spends as a neutron star and quark star, respectively.
   In our model the magnetic field during the neutron star era ($\tau_{\rm NS}$)   is
   $\sim 10^{12}$ G thus spin-down is  slow during this period. 
    The  delay time, $\tau_{\rm NS}$,  is defined by the time to reach deconfinement (Staff et al. 2006)
  plus a possible nucleation delay (Bombaci et al. 2004).
  However, after the QN the object's magnetic field is strongly magnified
   leading to a fast spin-down. 
  The result is that the spin-down age, $\tau_{\rm SD}$, is
  given by $\tau_{\rm SD}=\tau_{\rm QS}$, which is   
    determined by vortex expulsion and associated magnetic
  field decay (Ouyed et al. 2004; Ouyed et al. 2006; Niebergal et al. 2006; Niebergal et al. 2007).
   As shown in Ouyed et al. 2007a, the resulting spin-down age is  $0.16 P/\dot{P} \le \tau_{\rm SD}\le 0.33 P/\dot{P}$. This is reduced with respect to the standard value $P/(2\dot{P})$ by a factor
   of $\simeq 1.5$-3. 
   
   From Table 1, as noted above, 4 objects have spin-down ages much
   less than the minimum associated SNR age. The delayed amplification
    of magnetic field can alleviate this problem, i.e. $3000{\rm yr} < \tau_{\rm NS} < 9000\ {\rm yrs}$
     and $200{\rm yr} < \tau_{\rm SD} < 2000\ {\rm yrs}$.
    The only object in Table 1 that has spin-down age reliably greater
   than SNR age is AXP1E1841$-$045: the reduced spin-down age  in our model becomes
   consistent with the SNR age. In this case $\tau_{\rm NS}<<\tau_{\rm QS}$,
    thus no delay is required.  
    
    We have argued above that standard spin-down age estimates do not represent
    true ages. This affects the birth rate estimate for SGRs/AXPs given above in two ways.
    Firstly, the spin-down era is shorter by a factor of $\sim$ 1.5-3 or an average
    of 2.25. Secondly,
     the time since SN explosion is lengthened by the time delay to magnetic
     field amplification.  In reality the delay time is different for each object
       depending on the neutron star's initial period, magnetic field, and mass.
        As an approximation, we carried out fits with new age estimates using equation (\ref{eq:one})
      with fixed $\tau_{\rm NS}$, and with  $\tau_{\rm QS} =  \tau_{\rm SD}/2.25$. 
     For $\tau_{\rm NS} =$ 200, 500, 1000, 3000 yrs, the resulting birth rates were 
     $1/(316\ {\rm yr}),1/(400\ {\rm yr}),1/(510\ {\rm yr}),1/(875\ {\rm yr})$, respectively. Thus
      our previous estimates of $1/(500\ {\rm yr})$ to  $1/(1000\ {\rm yr})$  is valid but the uncertainty is increased.    This does not affect the main conclusion that about $1/5$ to $1/10$ of all core-collapse
      SN result in AXPs/SGRs.

 \subsection{AXPs 1E2259$+$586 and 4U0142$+$615}
 
 For these objects  we found  that their X-ray luminosity 
  is determined by accretion from a torus\footnote{The QN ejects the neutron star crust which forms a highly
      degenerate Keplerian torus for  rapidly spinning objects (Ouyed et al. 2007b).
      The  torus densities are representative of 
      neutron star crust matter  and can
      easily survive the strong radiation.} (Ouyed et al. 2007b). Thus their age
   as quark stars is $\tau_{\rm QS}\ne \tau_{\rm SD}$. This may explain
    the absurdly high spin-down age for 1E2259$+$586
     compared to its associated SNR age.
      We have previously argued that the same situation applies to 4U0142$+$615 (see Figure 1
      in Ouyed et al. 2007a).
      
      The estimate of $\tau_{\rm QS}$ is the time
      it takes to consume the torus or, $ \tau_{\rm QS}\sim m_{\rm t}/\dot{m}_{\rm t, q}$.
       The continuous (i.e. quiescent phase) accretion rate, $\dot{m}_{\rm t, q}$ is given 
        by eq.(20) in Ouyed et al. 2007b. We find
       \begin{equation}
       \tau_{\rm QS} \simeq 16000\ {\rm yrs}\ \frac{m_{\rm t,-7} M_{\rm QS,1.4}^{4}\mu_{\rm q,3.3}^6}{\eta_{0.1}^3R_{\rm t,25}^6}\ ,
       \end{equation}
        where: $m_{\rm t,-7}$ and $R_{\rm t,25}$ are the mass and radius
        of the torus in units of $10^{-7}M_{\odot}$ and 25 km, respectively;
          $M_{\rm QS,1.4}$ is the mass of the quark star in units if $1.4M_{\odot}$;
           $\mu_{\rm q}\sim 3.3$ is the mean molecular weight of the torus atmosphere, 
           and $\eta_{0.1}$ is the accretion efficiency in units of  $0.1$.    
  
\section{Conclusion}

Our study of  the birth rate of AXPs and SGRS  and their
 associated SNRs suggest that  about $1/5$ to $1/10$ of  all
   core-collapse SN lead to AXPs/SGRs.  
       These high rates suggest that all massive stars (greater
    than $ M_{\rm low}$) give rise to remnants with magnetar-like fields.
    
    This raises these issues: (i) 
    how  do all progenitors  with $M\ge M_{\rm low}$ generate
 $>10^{14}$ G fields in their compact remnants?; (ii)  why is there a dichotomy 
  in magnetic field strength between
  compact remnants from progenitors with mass greater than $M_{\rm low}$ (i.e.
   $B\sim 10^{14}$ G)
   and those with mass less  than $\sim M_{\rm low}$ ($B\sim 10^{12}$ G);
        (iii) why are all AXPs/SGRs isolated while many progenitors with $M>M_{\rm low}$ 
     are in binaries?

     In this study, we introduce the notion of delayed magnetic
     field amplification to resolve these issues.  
     We propose that  neutron stars from  progenitor masses $M> 9M_{\odot}$ 
       are born  with normal ($\sim 10^{12}$ G) magnetic
    fields. A neutron star from  a progenitor with an approximate
    mass range $M_{\rm low}< M <60 M_{\odot}$ 
     will experience an explosive transition to  a quark star (the QN) in which its magnetic
     field  is  amplified to $10^{14}$-$10^{15}$ G by color ferromagnetism (Iwazaki 2005).
      The second explosion (QN) and related mass loss helps to reduce
       the  surviving compact binary fraction  thus explaining the state
       of isolation of AXPs/SGRs.    
      The transition occurs with a delay of several hundred years (Staff et al. 2006).       
      This delayed amplification
        alleviates many  difficulties in interpreting simultaneously the high birth rate and high 
        magnetic fields of AXPs/SGRs and their link to massive stars.


\begin{acknowledgements}
This research is supported by grants from the Natural Science and
Engineering Research Council of Canada (NSERC).
\end{acknowledgements}





\begin{thebibliography}{}

\bibitem[Alpar(2001)]{2001ApJ...554.1245A} Alpar, M.~A.\ 2001, ApJ, 554, 
1245

\bibitem[]{} Berti, E., \& Stergioulas, N.\ 2004, MNRAS, 350, 1416

\bibitem[]{469} Bombaci, I., Parenti, I., \& Vidana, I. 2004, ApJ, 614, 314

\bibitem[Camilo et al.(2007)]{2007ApJ...666L..93C} Camilo, F.,  et al.\ 2007, ApJ, 666, L93 

\bibitem[Chatterjee et al.(2000)]{2000ApJ...534..373C} Chatterjee, P., 
Hernquist, L., \& Narayan, R.\ 2000, ApJ, 534, 373

\bibitem[]{476} Dar, A., \&  De R\'ujula, A.  2000, in 
  Results and Perspectives in Particle Physics (Ed. Mario 
Greco) Vol. 17, 13 

\bibitem[Ertan et al.(2006)]{2006ApJ...640..435E} Ertan, {\"U}., G{\"o}{\u 
g}{\"u}{\c s}, E., \& Alpar, M.~A.\ 2006, ApJ, 640, 435

\bibitem[Ferrario \& Wickramasinghe(2006)]{2006MNRAS.367.1323F} Ferrario, 
L., \& Wickramasinghe, D.\ 2006, MNRAS, 367, 1323

\bibitem[Gaensler et al.(1999)]{1999ApJ...526L..37G} Gaensler, B.~M., 
Gotthelf, E.~V., \& Vasisht, G.\ 1999, ApJ, 526, L37

\bibitem[Gavriil et al.(2002)]{2002Natur.419..142G} Gavriil, F.~P., Kaspi, 
V.~M., \& Woods, P.~M.\ 2002, Nature, 419, 142

\bibitem[Gill \& Heyl(2007)]{2007MNRAS.381...52G} Gill, R., \& Heyl, J.\ 2007, MNRAS, 381, 52

\bibitem[Gonzalez et al.(2007)]{2007arXiv0708.2756G} Gonzalez, M.~E., et al.\ 2007, ArXiv 
e-prints, 708, arXiv:0708.2756

\bibitem[Heger et al.(2005)]{2005ApJ...626..350H} Heger, A., Woosley, 
S.~E., \& Spruit, H.~C.\ 2005, ApJ, 626, 350

\bibitem[]{} Iwazaki, A.  et al. 2005,  Phys. Rev. D, 71, 034014

\bibitem[]{500} Iwazaki, A. 2005, Phys. Rev. D., 72, 114003

\bibitem[Ker{\"a}nen \& Ouyed(2003)]{2003A&A...407L..51K} Ker{\"a}nen, P., 
\& Ouyed, R.\ 2003, A\&A, 407, L51

\bibitem[Ker{\"a}nen et al.(2005)]{2005ApJ...618..485K} Ker{\"a}nen, P., 
Ouyed, R., \& Jaikumar, P.\ 2005, ApJ, 618, 485 

\bibitem[Leahy \& Ouyed(2007)]{2007arXiv0708.1787L} Leahy, D., \& Ouyed, 
R.\ 2007, ArXiv e-prints, 708, arXiv:0708.1787

\bibitem[Lyne et al.(1996)]{1996Natur.381..497L} Lyne, A.~G., Pritchard, 
R.~S., Graham-Smith, F., \& Camilo, F.\ 1996, Nature, 381, 497

\bibitem[Marsden et al.(2001)]{2001ApJ...550..397M} Marsden, D.,  et al.\ 2001, ApJ, 550, 
397

\bibitem[Manchester et al.(2005)]{2005AJ....129.1993M} Manchester, R.~N.,  et al.\ 2005, AJ, 129, 1993

\bibitem[]{} M\'esz\'aros, P., ``High-energy Radiation from Magnetized Neutron Stars" (Univ. of Chicago Press, 1992)

\bibitem[Niebergal et al.(2006)]{2006ApJ...646L..17N} Niebergal, B., Ouyed, 
R., \& Leahy, D.\ 2006, ApJ, 646, L17

\bibitem[Niebergal et al.(2007)]{2007arXiv0709.1492N} Niebergal, B., Ouyed, 
R., \& Leahy, D.\ 2007, ArXiv e-prints, 709, arXiv:0709.1492

\bibitem[Ouyed(2002)]{ouyed021} Ouyed, R., Dey, J., \& Dey, M. 2002, A\&A, 390, L39

\bibitem[Ouyed et al.(2004)]{2004A&A...420.1025O} Ouyed, R., et al.\ 2004, A\&A, 420, 1025 

\bibitem[Ouyed et al.(2006)]{2006ApJ...653..558O} Ouyed, R., et al.\ 2006, ApJ, 653, 558

\bibitem[Ouyed et al.(2007a)]{2007A&A...473..357O} Ouyed, R., Leahy, D., \& 
Niebergal, B.\ 2007a, A\&A, 473, 357

\bibitem[Ouyed et al.(2007b)]{2007b} Ouyed, R., Leahy, D., \& 
Niebergal, B.\ 2007b, A\&A, in Press [arXiv:astro-ph/0611133]

\bibitem[Ouyed et al.(2007)]{2007arXiv0705.1240O} Ouyed, R.,  et al.\ 2007, Submitted to ApJ [astro-ph/0705.1240]

\bibitem[Petit et al.(2007)]{2007arXiv0709.4526P} Petit, V., et al.\ 2007, ArXiv e-prints, 709, arXiv:0709.4526

\bibitem[]{} Press, W. H. et al. 1989, ``Numerical Recipes" (Cambridge University Press)

\bibitem[]{}  Savvidy, G. K. 1977,  Phys. Lett. 71 B, 113

\bibitem[Staff et al.(2006a)]{2006ApJ...645L.145S} Staff, J.~E., Ouyed, R., 
\& Jaikumar, P.\ 2006, ApJ, 645, L145

\bibitem[Thompson \& Duncan(1993)]{1993ApJ...408..194T} Thompson, C., \& 
Duncan, R.~C.\ 1993, ApJ, 408, 194

\bibitem[Vink \& Kuiper(2006)]{2006MNRAS.370L..14V} Vink, J., \& Kuiper, 
L.\ 2006, MNRAS, 370, L14 

\bibitem[Wang et al.(2006)]{2006Natur.440..772W} Wang, Z., Chakrabarty, D., 
\& Kaplan, D.~L.\ 2006, Nature, 440, 772

\end{thebibliography}
\end{document}